\documentstyle[aasms4,epsf]{article}
\def\kms{\rm{km \ s^{-1}}}
\def\etal{\rm{et al. }}
\def\deg{^\circ }
\def\solar{_\odot }
\tighten
\begin{document}
\title{Ongoing Gas Stripping in the Virgo Cluster Spiral NGC~4522}
\author {Jeffrey D. P. Kenney\altaffilmark{1}
\& Rebecca A. Koopmann \altaffilmark{1,2}}
\authoremail{kenney@astro.yale.edu, koopmann@astro.yale.edu}
\altaffiltext{1}{Yale University Astronomy Department, New Haven, CT
06520-8101}
\altaffiltext{2}{Union College, Department of Physics, Schenectady, NY 12308}
\abstract{The Virgo cluster galaxy NGC~4522 is one of the best spiral
candidates for ICM-ISM  stripping in action.
Optical broadband and H$\alpha$ images from the WIYN telescope
of the highly inclined  galaxy reveal a relatively undisturbed
stellar disk and a peculiar distribution of H$\alpha$ emission.
Ten percent of the H$\alpha$ emission arises from extraplanar HII regions 
which appear to lie within filamentary structures $\geq$3 kpc long above
one 
side of the disk. The filaments emerge from the outer edge of a disk of
bright H$\alpha$ emission which is abruptly truncated beyond 0.35R$_{25}$. 
Together the truncated H$\alpha$ disk and extraplanar H$\alpha$ filaments
are reminiscent of a bow shock morphology,
which strongly suggests that the interstellar medium (ISM) of NGC~4522 is
being stripped by the gas pressure of the intracluster medium (ICM).
The galaxy has a line-of-sight velocity of $\simeq$1300 $\kms$ with
respect to
the mean Virgo cluster velocity, and thus is expected to experience a
strong interaction with the intracluster gas.
The existence of HII regions apparently located above the disk plane
suggests that star formation is occuring in the stripped gas, and that
newly formed stars will enter the galaxy halo and/or intracluster space.
The absence of HII regions in the disk beyond 0.35R$_{25}$, and the
existence of HII regions in the stripped gas suggest that even
molecular gas has been effectively removed from the disk of the galaxy.
}

\section {Introduction}

Among the many types of galaxy interactions which are
posited to occur in clusters,
one of the most important may be
the stripping of gas from the interstellar media (ISM) of galaxies 
due to interactions with the gas in the intracluster medium (ICM)
(Gunn \& Gott 1972).
This process of ICM-ISM stripping is likely to
significantly affect the  morphology and evolution
of cluster spiral galaxies, and may be one of the factors which explains
the morphology-density relationship (Dressler \etal 1997).
It remains the best explanation for several observations:
the large number of HI deficient
spiral galaxies observed in clusters (Giovanelli \& Haynes 1983), 
spiral galaxies with truncated HI disks yet 
relatively undisturbed stellar disks (Warmels 1988; Cayatte \etal 1990)
and cluster spirals with vigorous star formation in their
central regions, but very little throughout most of their outer disks
(van den Bergh \etal 1990; Koopmann \& Kenney 1998a,b).
Yet despite widespread indirect evidence for the ICM-ISM stripping of
cluster
spiral galaxies, there are few clear examples of gas actively being
stripped from the disks of spirals by the ICM. 

Instead, the best cases so far for ongoing ICM-ISM 
stripping are elliptical galaxies.
Evidence seems quite strong that stripping 
of the outer hot halo gas occurs in
the Virgo ellipticals 
NGC~4406 (M86; Forman \etal 1979; White \etal 1991),
NGC~4472 (M49; Irwin \& Sarazin 1996) and perhaps
NGC~4636 (Trinchieri \etal 1994).
The ease and consequences of ICM-ISM stripping
for ellipticals are different from those for spirals,
primarily because the hot ISM in ellipticals is a halo component,
which has a much lower density than the disk gas of spirals. 
This hot gas does not efficiently form stars, so its 
loss does not greatly effect the elliptical's future evolution.
However, the loss of star-forming disk gas from spirals would have
a greater evolutionary consequence, perhaps
accelerating their evolution to a lenticular morphology.

Perhaps the best spiral/irregular candidates for ongoing ICM-ISM stripping
are three galaxies (97073, 97079, 97087) with peculiar H$\alpha$ 
and non-thermal radio continuum properties located in the outskirts
of the cluster A1367 (Gavazzi \etal 1995). These galaxies have
radio continuum emission extended well beyond the optical disks,
with head-tail morphologies, and strongly enhanced star formation rates,
including bright HII regions, which in at least one case
lies in an arc in the outer disk of the galaxy. 
Among more nearby spiral galaxies, there are several with peculiarities 
that have been attributed to ongoing ISM-ICM stripping, although for 
most of these galaxies the evidence is less compelling.
The group galaxy NGC~7421 has a bow-shock shaped HI morphology,
suggestive of an ISM-ICM interaction (Ryder \etal 1997).
The cluster or group spiral galaxies NGC~1961, NGC~2276, NGC~3312,
NGC~4438, NGC~4654, and NGC~5291 
(respectively Shostak \etal 1982; Mulchaey \etal 1993;
Gallagher 1978; Kotanyi \etal 1983; Phookun \& Mundy 1995; 
Malphrus \etal 1997)
have all been cited as possible examples of ICM-ISM stripping in action.
However, for each of them, a tidal or other
origin for their peculiarities
has also been suggested (Pence \& Rots 1997; Gruendl \etal 1993;
McMahon \etal 1992; Kenney \etal 1995; Cayatte \etal 1990; 
Malphrus \etal 1997).
In each of these systems, the stellar as well as the gaseous 
distribution appears to be disturbed, thus in none of these
cases is it clear that ICM-ISM stripping is responsible for the
galaxy peculiarity.
In contrast to the many spiral galaxies which are widely recognized as 
being in various stages of gravitational interactions and mergers,
there are no generally recognized ``smoking guns'' for ICM-ISM stripping in 
spirals.

In this paper, we present broadband BVR and narrowband H$\alpha$ images
of
the relatively obscure, highly inclined Virgo cluster spiral galaxy
NGC~4522.
Its relatively normal stellar disk and selectively disturbed ISM 
strongly suggest that its ISM is actively being stripped by
an ICM-ISM interaction, and
we consider NGC~4522 an excellent nearby spiral candidate for ongoing 
ICM-ISM stripping.

\section {The Galaxy NGC~4522}

The optical peculiarities of NGC~4522 are reflected in its classification
of SBcd:(sp) (RC3) and Sc/Sb: (Binggeli, Sandage, \& Tammann 1985). 
The uncertain Sc/Sb Hubble classification probably results
from the small bulge and the relatively high star formation rate in the 
inner region (like Sc), in combination with the lack of star formation 
in the outer
disk (more like Sb). The galaxy is highly inclined (i=75$\pm$5$\deg$),
making it possible to see extraplanar features.
The ``spindle'' designation from the RC3 probably arises from the
extraplanar dust and HII regions, which are discussed below.
A photograph in Sandage \& Bedke (1994) shows an ``arm fragment''
emerging from the disk, which we show to be composed primarily 
of HII regions.
It is significant that all the known peculiarities in NGC~4522 are
associated with dust, gas, or HII regions, and not older stars.

Although it was not mapped in HI in either the
Westerbork (Warmels 1988) or VLA (Cayatte \etal 1990) 
Virgo cluster HI surveys, it has been detected in HI with
the Arecibo telescope (Helou \etal 1984).
It has an HI deficiency of 0.6,
meaning that NGC4522 contains only $\sim$1/4 the amount
of atomic gas that a normal late type spiral of its size would contain
(Giovanelli \& Haynes 1983; Kenney \& Young 1989).
It is 3.3$\deg$ from M87, which is at the center of the main galaxy 
concentration in Virgo, 
and only 1.5$\deg$ from M49, which is at center of subcluster B
(Binggeli, Popescu, \& Tammann 1993).
Binggeli, Sandage, \& Tammann (1985) consider it a member of the Virgo
cluster,
and Yasuda \etal (1997) find a Tully-Fisher distance 
consistent with Virgo cluster membership.
This part of the Virgo cluster contains many HI-deficient
galaxies (Haynes \& Giovanelli 1986),
as well as x-ray emission from hot intracluster gas (Bohringer \etal
1994).
The ROSAT map shows weak extended x-ray emission at the 
projected location of NGC~4522 (B\"{o}hringer \etal 1994), 
although NGC~4522 itself does not appear to be
a strong localized source of x-ray emission
(Fabbiano \etal 1992; B\"{o}hringer \etal 1994; Snowden 1997).
A compilation of galaxy properties is given in Table 1.

\section {Observations}

Exposures of NGC~4522 were taken in BVR and narrowband H$\alpha$+[N II]
filters on the 3.5 m WIYN telescope$\footnote{
The WIYN Observatory is a joint facility of the University of Wisconsin-Madison, 
Indiana University, Yale University, and the National Optical Astronomy 
Observatories.}$
at KPNO in April 1997.
A  2048$\times$2048 S2KB CCD with a plate scale of 0.2$''$/pixel was used, 
giving a field of view of 6.8$'$ (31 kpc). 
The narrowband filter had a bandwidth of 70\AA\ centered on 6625\AA ,
and included the redshifted H$\alpha$ and [N II] lines.
Integration times totalled 10 minutes for B and V, 20 minutes for R, 
and 30 minutes for H$\alpha$+[N II], and were divided up into
3 or more exposures per filter.
The seeing was 1.1$''$. 
We used the IRAF package in a standard way
to bias-subtract, flat-field,
register, and combine the images for each filter.
Cosmic rays were removed using the pixel rejection routines in the
combine task.
After sky subtraction, the R-band image was scaled and 
then subtracted from the narrowband filter image
to obtain a continuum-free H$\alpha$+[N II] 
(hereafter referred to as an H$\alpha$) image. 

\section {The Peculiar H$\alpha$ Morphology of NGC~4522 and Possible
Explanations}

The B-band and H$\alpha$ images
of NGC~4522 are shown in Figures 1 and 2, and contour maps
produced from R-band and H$\alpha$ images smoothed to 3$''$ resolution
are shown in Figure 3.
There is no evidence from any of the broadband images or contour maps
of any significant peculiarity in the distribution of older stars.
The presence of dust and star-forming regions in the inner galaxy
strongly affects the observed light distribution in the
central
r$\simeq$60$''$, but there is less dust and
little or no massive star formation beyond this radius.
R-band surface photometry shows that the stellar light distribution
beyond r=20$''$ and out to at least r=170$''$ is well fit by an 
exponential with a scale length
of 30$\pm$3$''$, and that the bulge contribution is very small
(Koopmann, Kenney, and Young 1998).

In contrast, the H$\alpha$ morphology is distinctly peculiar.
Ten percent of the total H$\alpha$ emission 
arises from a one-sided, extraplanar distribution,
predominantly organized into filaments which extend more than 35$''$ = 3
kpc
from the outer edge of a truncated  H$\alpha$ disk.
The extraplanar H$\alpha$ emission arises from both HII regions (Fig 2), 
and diffuse emission (Fig.3a). 
The H$\alpha$ morphology is reminiscent of a bow shock, and,
when coupled with a normal stellar disk, strongly
indicates that the ISM of NGC~4522 is being 
selectively disturbed.

The outermost bright HII complex in the southwestern part of the disk,
at a projected radius of 40$''$=3.1 kpc,
has an interesting bubble morphology in H$\alpha$, as shown in Fig. 2.
This relatively uniform surface brightness shell 
with a diameter of 5$''$=400 pc
extends over 180$\deg$, 
and appears to emerge from
two bright HII regions at its base
The regular morphology of the shell 
suggests that a wind from supernovae
and OB stars
is expanding into a relatively uniform medium, perhaps the ICM.

What internal processes could selectively disturb gas in galaxies?
Starbursts and cooling flows produce
distinctive ionized gas morphologies which clearly differ from that
in NGC~4522.
Intense starbursts have outflowing gas with double bubble (Jogee \etal
1998)
or biconical (Heckman \etal 1990) extraplanar
H$\alpha$ distributions. Extraplanar gas is generally produced on both
sides of the disk, and
originates from a location which is near the galaxy center and
inside a region of active star formation. 
``Cooling flow'' galaxies often exhibit filamentary H$\alpha$ 
morphologies which extend from all sides of the galaxy center 
(e.g. Lynds 1970).
Some edge-on spiral galaxies contain extraplanar H$\alpha$ filaments or 
diffuse emission, probably caused by massive star formation within the
disk
(Rand, Kulkarni, \& Hester 1990; Pildis, Bregman, \& Schombert 1994; Rand
1997).
These filaments are generally smaller and less luminous than those in NGC
4522,
and are not selectively located near the outer edge of the star-forming
disk,
as are those in NGC~4522.

Galaxy-galaxy gravitational interactions are also unlikely to explain the
morphology of NGC 4522. While gravitational interactions affect gas
differently 
than stars, both stars and gas are disturbed (e.g., Barnes \& Hernquist
1992).
Given the degree of morphological peculiarities in H$\alpha$,
one would expect to see some disturbance in the stars of NGC 4522, such as
a tail or warp.
For example, while the peculiar Virgo cluster galaxy NGC~4438 has a 
one-sided set of extraplanar ionized gas filaments (Kenney \etal 1995)
somewhat like those in NGC~4522, the stellar disk of NGC~4438
is highly disturbed, indicating that NGC~4438 has experienced
a high-velocity galaxy-galaxy interaction(Kenney \etal 1995;
Moore \etal 1996).

The distinctive H$\alpha$ morphology of NGC~4522, together with the 
normal-appearing stellar disk, not only seems inconsistent with 
vigorous disk star formation, a central starburst, a cooling flow, 
or a galaxy-galaxy interaction, but
is what is generally expected for ICM-ISM stripping.
Moreover, the extraplanar optical emission in 
edge-on star-forming galaxies, starbursts, cooling flows,
and NGC~4438 have line ratios and morphologies 
indicating shock-excited, diffuse gas 
(Heckman \etal 1989; Heckman \etal 1990; Kenney \etal 1995; Rand 1997), 
rather than HII regions, as is observed in NGC~4522.
We therefore propose that the ISM of NGC~4522 is
being stripped by the gas pressure of the intracluster medium (ICM).

Given the large number of HI-deficient galaxies in clusters,
it is reasonable to wonder why NGC~4522 should be
such a good case for observing ongoing stripping.
We believe that both intrinsic properties and favorable viewing
circumstances are responsible.
The galaxy is especially susceptible to stripping, due to its high
speed and low mass.
Its radial velocity of 2330 $\kms$ is on the high end of the 
distribution for Virgo galaxies, implying that it has a
velocity of at least 1280 $\kms$ with respect to the mean cluster velocity
of 1050 $\kms$ (Binggeli, Popescu, \& Tammann 1993).
There must also be a significant velocity component in the plane of the 
sky 
to cause the extraplanar filaments in this nearly edge-on galaxy.
The high velocity ensures a strong ICM-ISM interaction, 
since ram pressure is proportional to the square of the velocity.
Long-slit H$\alpha$ spectroscopy along the disk plane shows a
fairly normal, rising rotation curve, with V$_{\rm rot}$=103 $\kms$
at a radius of 42$''$ (Rubin, Waterman \& Kenney 1998).
The fairly low mass of the galaxy indicated by this rotational speed
implies that the ISM of NGC~4522 will be less tightly bound to the galaxy
than in brighter, more massive spirals.
NGC~4522 also has a favorable viewing angle.
Since it is close to edge-on, it is easier to see extraplanar gas.
In addition, we are likely to be viewing it at a particularly favorable
time,
perhaps on its first passage through the dense part of the ICM.

\section {Star Formation in NGC~4522}

There has been much discussion in the literature of galaxies 
with star
formation rates enhanced due to triggering by ICM-ISM interactions
(Gavazzi \etal 1995; Dressler \& Gunn 1983).
The star formation properties of NGC~4522 can be 
assessed with Figure 4, which shows
normalized H$\alpha$ luminosities versus
central light concentrations for samples of Virgo cluster (bottom) and 
isolated (top) spiral galaxies.
In this figure, which is adapted from Koopmann \& Kenney (1998a),
the concentration parameter C30 is the flux ratio of the R-band light
within 0.3R$_{24}$ to that within R$_{24}$, and is an objective tracer
of the stellar bulge-to-disk ratio.
The H$\alpha$ luminosity normalized by the the R-band luminosity
is a measure of the present-day
massive star formation rate divided by the past-average star formation
rate.
Figures 4a and b show the {\it global} H$\alpha$ to R luminosity ratio,
Figures 4c and d show this ratio for the {\it inner disk},
i.e., within 0.3R$_{24}$, and 
Figures 4e and f show this ratio for the {\it outer disk},
i.e., the annulus from 0.3-1.0R$_{24}$.
The dotted lines in each panel bound the values found for the isolated
spirals 
(but not S0's).

Figure 4b shows that the global L(H$\alpha$)/L(R) for NGC~4522 
is slightly below average, compared with
both isolated and Virgo spirals of the same central light concentration.
Since the H$\alpha$ luminosities are not corrected for internal
extinction,
and this correction should be higher for highly inclined galaxies like
NGC~4522, the true global NMSFR for NGC~4522 
is probably close to average.
Its modest infrared-to-optical luminosity ratio
L$_{\rm FIR}$/L$_{\rm opt}$=0.5 (Table 1) suggests that the extinction
correction is not too large.
Figure 4d shows that the \it central \rm L(H$\alpha$)/L(R) for NGC~4522 
is somewhat above average, compared with both isolated and Virgo spirals.
Correcting for extinction would presumably elevate NGC~4522 even higher
above the average.
Thus it is plausible that the star formation rate in the center of
NGC~4522 has been modestly enhanced by a factor of $\sim$2 by an ICM-ISM
interaction.
The other Virgo galaxies with similar or higher central star formation
rates show evidence for some kind of interaction (Koopmann \& Kenney
1998b).
Figure 4f shows that the outer disk L(H$\alpha$)/L(R) for
NGC~4522 
is well below average.
Thus a lack of star formation in the outer disk, together with moderately
enhanced star formation in the inner disk, combine to produce an
average global NMSFR.

Approximately 10\% of the global H$\alpha$ flux arises from HII regions
which appear to be located above the disk plane, in gas 
presumably stripped from the disk. 
This corresponds to a total extraplanar star formation rate of 
0.01 M$\solar$ yr$^{-1}$, according to the assumptions of Kennicutt
(1983).
The stars which form in this outflowing gas will
enter the galaxy halo or intracluster space, depending on whether the
gas is bound to the galaxy at the time when the stars form.
This may be a source of newly formed intracluster stars, which is relevant
in
light of the recent discovery of intracluster stars in the Virgo cluster
(Ferguson, Tanvir \& von Hippel 1998).

This star formation rate is probably much lower than the gas stripping
rate.
NGC 4522 has an HI mass of 0.4$\times$10$^9$ M$\solar$,
compared to the average value of 1.4$\times$10$^9$ M$\solar$
for an isolated spiral galaxy of the same optical diameter
(Giovanelli \& Haynes 1983), suggesting that
$\sim$10$^9$ M$\solar$ of gas been stripped from the galaxy.
If this gas is stripped over a cluster crossing time of 10$^9$ yrs,
this corresponds to an average stripping rate of 1 M$\solar$ yr$^{-1}$.
This is significantly larger than both the global SFR of 
0.1 M$\solar$ yr$^{-1}$, and the extraplanar SFR of 0.01 M$\solar$
yr$^{-1}$.
Thus most of the stripped gas is not converted into stars by the 
ICM-ISM interaction.
While in galaxy disks gas can continually and quickly cycle through
phases and return to dense star-forming gas clouds,
this probably doesn't happen in the stripped gas. 
Instead, most of the stripped gas likely escapes the galaxy, where 
it is heated by and then joins the
hot intracluster medium, enriching its abundance of heavy elements.
The effect of this ICM-ISM interaction on galaxy evolution is that
stellar diskbuilding is truncated  in time  in the
outer disk, but is normal or perhaps even somewhat accelerated 
in the inner disk, if the inner star formation rate is enhanced.
Integrated over time, the
outer disks never achieve their maximum stellar mass surface density,
whereas inner disks are little affected.

\section {Is Molecular Gas Stripped?}

The absence of HII regions in the outer disk, and the existence of HII
regions
in the extraplanar gaseous filaments suggests that even molecular gas
has been removed from the outer disk as a consequence of stripping.
According to the ram pressure stripping criterion of Gunn \& Gott (1972),
gas is stripped if the ram pressure $\rho$v$^2$ exceeds the gravitational
force per unit area G$\sigma _{\rm gas}$ $\sigma _{\rm tot}$/2$\pi$
binding gas to the disk.
Thus gas at a surface density of $\sigma _{\rm gas}$=20 M$_{\sun}$
pc$^{-2}$
would be stripped from the disk of NGC 4522 at r=5 kpc,
if the galaxy is moving at v=1500 km s$^{-1}$ wrt the ICM,
and the ICM density is $\rho$=10$^{-4}$ cm$^{-3}$ 
(Fabricant \& Gorenstein 1983).
This assumes a spherical galaxy mass distribution, and a rotation speed 
(in the plane of the galaxy) of 103 km s$^{-1}$ at r=5 kpc (Rubin \etal
1998).

The value of $\sigma _{\rm gas}$=20 M$_{\sun}$ pc$^{-2}$
resulting from this simple calculation
is nearly an order of magnitude lower than the
gas (H$_2$+He) surface densities of 170 M$_{\sun }$ pc$^{-2}$ found for
inner disk (1$\leq$R$\leq$8 kpc) Milky Way GMCs (Larson 1981; Solomon
\etal 1987), but
within a factor of 4 of the value of 80 M$_{\sun }$ pc$^{-2}$ 
found for the outer disk (8$\leq$R$\leq$24 kpc)
Milky Way clouds described by Heyer \etal (1998).
Molecular cloud surface densities are unlikely to be a universal constant,
since they are clearly much higher than 170 M$_{\sun }$ pc$^{-2}$ 
in the centers of many galaxies (Kenney \etal 1993; Scoville \etal 1997),
and may be lower than this in the outer disks of galaxies.
We cannot say whether the stripped gas surface density is as low as 
20 M$_{\sun}$ pc$^{-2}$ in NGC~4522, but other effects may
allow the effective stripped surface density to be higher than this.

One possibile effect is that NGC~4522 is currently passing
through a region with above average ICM density,
although neither Einstein nor ROSAT maps show 
any evidence for enhanced x-ray emission near NGC~4522 
(Fabbiano \etal 1991; Snowden 1997). 

It might be possible to strip clouds with surface densitities of 
100-200 M$_{\sun }$ pc$^{-2}$  
with the same ram pressure,
if the effective area for stripping were much larger than an individual
molecular cloud, so that the appropriate gas surface density in 
the ram pressure stripping equation
were correspondingly lower. This might occur, for example, if magnetic
fields
coupled GMCs to the surrounding intercloud medium. 
If the effective length scale
for stripping were 1 kpc instead of 100 pc, then the strippable
gas surface density might be an order of magnitude lower.

It also may be possible to strip the outer disk of all gas and star
formation, 
even without directly stripping GMCs.
GMCs may be short-lived, and the gas may be
stripped when the gas is in a non-molecular, lower density phase.
GMCs which form massive stars are likely to be short-lived, with lifetimes 
of $\sim$10$^7$ yrs (Elmegreen 1991; Larson 1994), due to the 
enormous energy input into the cloud from massive stars and supernovae.
If most of the GMC mass is cycled through a low surface density phase
over timescales shorter than a galaxy rotation period
(5$\times$10$^7$ yrs at r=5 kpc in NGC~4522), then most of the ISM at a 
given radius can be stripped, if the ram pressure stripping equation 
holds for the minimum surface density through which most of the ISM
cycles.
While gas cycling is surely an important factor affecting the response
of a galaxy's ISM to an interaction with the ICM, this cannot be the
sole factor in the stripping of NGC~4522's star-forming clouds,
unless the HII regions which exist in the extraplanar gas formed from
gas which became dense after being stripped.

Although most HI-deficient Virgo cluster (Kenney \& Young 1989; Kenney
1990) and the 3 peculiar A1367 cluster spirals (Gavazzi \etal 1995)
have relatively normal CO luminosities (Boselli \etal 1994) , only the outer disk is 
stripped of gas in spiral galaxies, meaning that
only a modest fraction of the total H$_2$ is typically 
lost due to stripping. 
Stripping only outer disk gas  changes the global CO luminosity of 
spirals by only a modest amount.
There are many post-stripped spirals in Virgo 
with relatively normal massive star formation rates in the inner disk,
and virtually no massive star formation in the outer disks
(Koopmann \& Kenney 1998b). 
Such galaxies could have modest
reductions in CO luminosity, without being noticeable.
The most severely stripped cluster galaxies could be classified as S0's,
and the CO normalcy of this heterogeneous group is unknown.

\section {Comparisons between NGC 4522 and Other Candidate Strippers}

NGC 4522 exhibits several interesting differences from 
other ICM-ISM stripping candidates, particularly the three candidates
in A1367. In NGC 4522, HII regions are
clearly located above the disk plane in the wake of the interaction. 
While the global SFR in NGC 4522 is normal to reduced, 
the three A1367 stripping candidates have strongly enhanced SFRs 
(Gavazzi \etal 1995). NGC 4522 is HI deficient, while the three A1367
galaxies have nearly normal HI content (Dickey \& Gavazzi 1991).
Their HI distributions seem to be off-center, although the
existing HI maps have low S/N (Dickey \& Gavazzi 1991). 
These galaxies, particularly 97073,
resemble the Virgo Cluster spiral NGC 4654 more closely than NGC 4522.
While NGC 4654 has an HI content and star formation rate 
typical for a late-type spiral,
its HI distribution has a head-tail
morphology, and on the opposite side of galaxy from the HI tail
there is a curved ridge of bright HII
regions forming the edge of its outer disk (Phookun \& Mundy 1995).
Perhaps the main difference between NGC~4654 and the three
A1367 galaxies is the ICM density, which is nearly an order of magnitude 
larger in A1367 (Gavazzi \etal 1995).
This may account for the strongly enhanced SFRs in the A1367 galaxies,
compared to the modest enhancements in the SFRs of NGC~4654 and other
Virgo galaxies.

NGC~4522 is already HI-deficient and has a shrunken star forming disk,
suggesting that stripping has been going on for a longer time in NGC~4522
than NGC 4654 and the  three A1367 galaxies.
There are a  number of Virgo spiral galaxies with truncated
star forming disks, which are likely to be 
undergoing or to have undergone ICM-ISM stripping:
NGC~4064, NGC~4380, NGC~4405, NGC~4413, NGC~4580, and IC~3392
(Koopmann \& Kenney 1998a,b).
NGC~4522 has a higher star formation rate than the other
truncated spirals identified in Koopmann \& Kenney (1998b), 
suggesting that it might be at an intermediate evolutionary phase 
in ICM stripping, more advanced than
the 3 A1367 galaxies and NGC~4654,
yet less advanced than the Virgo
spirals with severely truncated star forming disks.

\section {Future Work}

While its optical morphology alone strongly suggests that NGC~4522 
is experiencing ICM-ISM stripping, HI mapping will clearly reveal
much more about this galaxy's interaction with its environment.
One question among many is whether there is
any evidence for a shock at the ICM-ISM interface.
There is no strong x-ray emission from 
NGC~4522 or any other non-Seyfert Virgo spiral, 
or from any of the 3 A1367 cluster spirals (Gavazzi \etal
1995), suggesting that 
strong x-ray emission is not produced by ICM-ISM stripping in spirals.

\section{ Acknowledgements}

We are grateful to Wendy Hughes and Orion Peck for processing some of the 
optical images, Mark Heyer and Bev Smith for helpful discussions, and
Steve Snowden for providing a ROSAT x-ray map prior to
publication. Thanks especially to the WIYN staff for their excellent 
support at the telescope. This work has been partially supported by 
NSF grant AST-9322779.

\newpage

\figcaption[kenney.fig1.ps]{
The WIYN B image 
 of NGC~4522 with a  6.5$'$ (30 kpc) 
field of view. (A color BVR image is available on-line.)
The stellar disk is relatively undisturbed.
The peculiarities are due to dust and HII regions towards the
NW which appear to be extraplanar.}

\figcaption[kenney.fig2.ps]{
The WIYN H$\alpha$ image of NGC~4522 with a  88$''$$\times$97$''$ (7.5 kpc) 
field of view.
Note strings of apparently extraplanar HII regions emerging
mostly from outer edge of coplanar star-forming disk.
Also note H$\alpha$ bubble 40$''$ SW of nucleus.}

\figcaption[kenney.fig3.ps]{
Contour and greyscale plots of NGC~4522, smoothed to 3$''$ resolution.
(a) H$\alpha$, showing truncated star-forming disk, and extraplanar HII 
regions.
(b) R-band, showing fairly regular elliptical isophotes at large radii,
suggesting that the stellar disk is undisturbed.
(c) H$\alpha$ grayscale on R-band contours, showing disturbed
H$\alpha$ distribution with respect to undisturbed stellar disk.
Images have been rotated by 57$\deg$ so that galaxy is oriented
horizontally.}

\figcaption[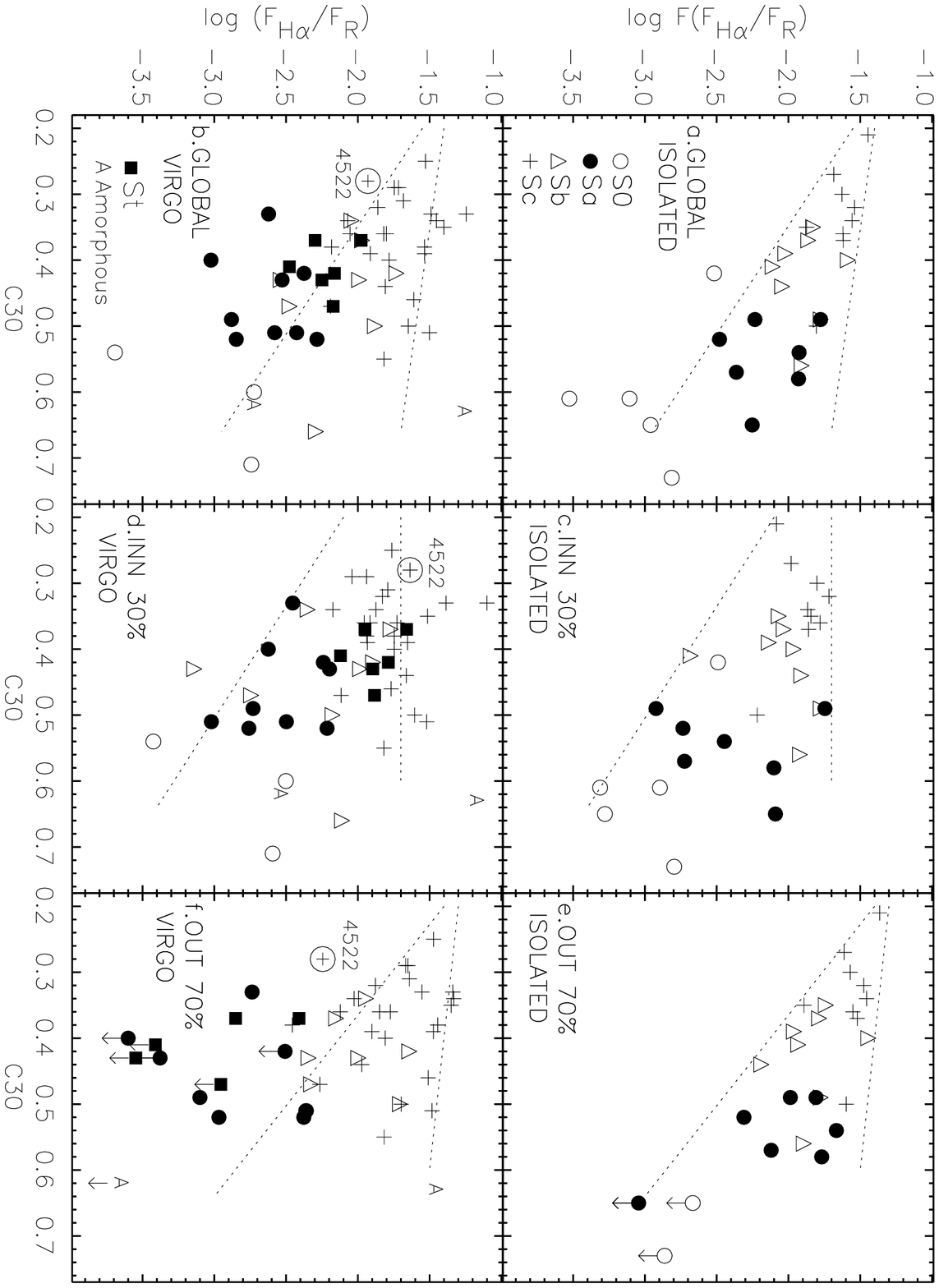]{
Central light concentration C30 versus \it global \rm normalized 
massive star formation rate
F(H$\alpha$)/F(R) for (a) 29 isolated spiral galaxies, and 
(b) 55 Virgo cluster spiral galaxies, from Koopmann \& Kenney (1998a).
The position of NGC~4522 is indicated. Dotted lines indicate the
approximate bounds of values for isolated spiral galaxies.
This diagram shows that the global normalized star formation rate
for NGC~4522 is slightly below average for spiral galaxies with its 
bulge-to-disk ratio.
(c, d)
Same as a and b, except that the \it inner \rm normalized 
star formation rate is shown,
i.e. the ordinate is the H$\alpha$ to R flux ratio 
within the central 30\% of R$_{24}$.
This diagram shows that the central normalized star formation rate
for NGC~4522 is slightly above average, compared with
isolated spiral galaxies of the same bulge-to-disk ratio.
(e, f)
Same as a and b, except that the \it outer disk \rm normalized 
star formation rate is shown,
i.e. the ordinate is the H$\alpha$ to R flux ratio in the
annulus r=0.3-1.0 R$_{24}$. For NGC 4522, the HII regions which 
seem to lie above the disk plane have not been included in the
outer disk total.
This diagram shows that the outer disk
normalized star formation rate
for NGC~4522 is well below average, compared with
isolated spiral galaxies of the same bulge-to-disk ratio.}

\end {document}